# Measurement of radiant spectrum for excess heat generation in NiCu and Ni thin film during hydrogen gas desorption [#i]

J. Kasagi [1], T. Itoh [1,2], Y. Shibasaki [3], Y. Iwamura [2]

[1] Research Center for Accelerator and Radioisotope Science, Tohoku University, Japan
[2] Yokohama City University, Japan
[3] Clean Planet Inc., Japan
Email: kasagi@lns.tohoku.ac.jp

**Abstract:**
Heat production in Ni-based thin films interacting with hydrogen gas was studied using a radiation calorimeter equipped with a composite photon-detector system. Measurements were performed on pure Ni foil and NiCu composite thin films. Electromagnetic radiation emitted from samples heated to approximately 1100 K in vacuum was recorded over the energy range of 0.05–2.5 eV. In this study, detailed spectra were obtained, particularly in the low-energy region below 0.5 eV. The peak of the thermal radiation spectrum was observed around 0.2 eV, deviating from that expected for simple gray-body radiation. Spectral variations depending on the presence or absence of hydrogen differed between the NiCu and pure Ni samples. The total radiated energy was determined directly from the measured radiation power, without relying on radiation models. The excess heat generated in the NiCu sample was estimated to be approximately 1.1 W. A comparable but slightly smaller excess heat generation was also observed in Ni. Long-term measurements of the NiCu sample demonstrated sustained excess heat production for at least 215 hours, with a decay time exceeding 2000 hours.

## 1. Introduction

Much work on the phenomenon of anomalously large heat generation in metal–hydrogen (or deuterium) systems has been reported since the work of Fleischmann and Pons [1]. They observed excess heat during electrolysis of heavy water with a Pd cathode and claimed that the heat produced far exceeded the enthalpy of any chemical reaction and suggested that the underlying cause was the DD fusion reaction, i.e., "cold fusion." Since then, various experimental approaches have been explored, including not only electrolysis but also gas loading, ion implantation, and other methods.

In terms of metallic materials, a wide variety of forms have been tested, such as Pd, Ni, and Cu in the form of bulk metal, fine particles, amorphous powders, and nanostructured composite thin films [2,3]. It is increasingly recognized that the anomalous heat generation may involve a process fundamentally different from conventional nuclear reactions. In particular, the emission of high-energy particles such as gamma rays and neutrons is rarely observed, with most of the released energy appearing as thermal output.

Since nanostructured composite metals have shown promise as heat-generating materials [4], we have been investigating anomalous heat generation phenomena in metal thin films for several years [5–7]. Our group has developed a method to induce large excess heat in composite NiCu thin films exposed to $H_2$ and $D_2$ gas. To obtain solid evidence for this heat production, we have established a radiation calorimetry technique. In our experiments, a thin metal film is placed in vacuum and heated to high temperatures [5,6], where most of the heat transfer from the sample occurs via electromagnetic radiation. Accurate determination of the total heat flow therefore requires measurement of the radiation spectrum over a wide energy range.

In earlier measurements, three spectroscopic instruments with different energy ranges were employed to reconstruct the full spectrum by combining their data. However, in the range below 0.5 eV, only the average intensity between 0.2 and 0.4 eV was recorded, leaving the spectrum incomplete. To address this limitation, we recently incorporated a Fourier-transform infrared (FT-IR) spectrometer, which covers a much broader wavelength region from the far-IR to the near-IR, corresponding to photon energies from 0.004 eV to 0.93 eV.

Obtaining the complete radiation spectrum, including the low-energy region below 0.5 eV, allows us to analyze the data without relying on the previously used gray-body radiation approximation, thereby enabling a more model-independent and quantitatively reliable determination of the generated energy.

## 2. Detection system for photon radiation from metal thin film; BREMS

Figure 1 shows a simplified cross-sectional view of the entire setup, seen from above. The detection system, which we call BREMS (Broad Range ElectroMagnetic wave detection System), consists of a newly introduced FT-IR spectrometer and four previously employed spectrometers. Each detector is directed toward a sample placed at the center of the vacuum chamber. Two thin-film samples are mounted on either side of a ceramic heater suspended from the chamber.

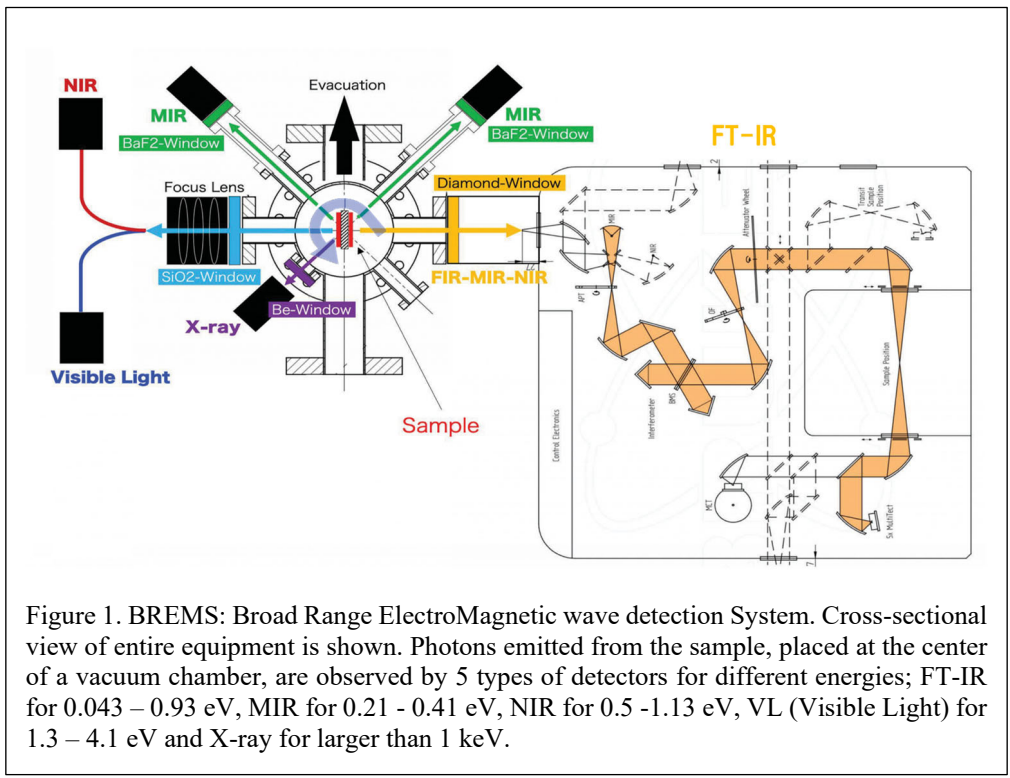

Figure 1. BREMS: Broad Range ElectroMagnetic wave detection System. Cross-sectional view of entire equipment is shown. Photons emitted from the sample, placed at the center of a vacuum chamber, are observed by 5 types of detectors for different energies; FT-IR for 0.043 – 0.93 eV, MIR for 0.21 - 0.41 eV, NIR for 0.5 -1.13 eV, VL (Visible Light) for 1.3 – 4.1 eV and X-ray for larger than 1 keV.

The newly installed spectrometer, referred to as FT-IR, is shown on the right side of Fig. 1. It is a Fourier transform spectrometer (INVENIO R, Bruker) that covers the energy range from 0.043 eV to 0.93 eV (wavelength 28.5 - 1.33 μm) in MIR mode and from 0.004 eV to 0.084 eV (wavelength 325 - 14.8 μm) in FIR mode. Photons emitted from the sample enter the FT-IR through a diamond window located 20 cm from the sample, passing through a 10-mm diameter aperture that selects parallel rays. To avoid absorption by $H_2O$ and $CO_2$ molecule, nitrogen gas supplied is continuously supplied inside the FT-IR, displacing the air.

On the opposite side of the camber, a SiO2 window transmits photons from the near-IR to the UV range. These photons pass through a focusing lens and are guided into an optical fiber, which is split between two spectrometers; NIR, an FT spectrometer (Hamamatsu C15511) covering 0.5 - 1.13 eV (1.5- 2.5 μm), and VL (Visible Light), a multichannel spectrometer (Hamamatsu C10027) covering 1.3 - 4.1 eV (0.3-0.9 μm).

For improved statistical accuracy, all three spectrometers output the average radiant power from multiple scans as a function of wavelength or wavenumber. The acquisition times are ~30 s for NIR, ~50 s for VL, and ~60 s for FT-IR. Photon collection is restricted to within a ~10-mm diameter around the sample center.

In addition to these two main windows, the vacuum chamber has four auxiliary windows, each at a 60-degree angle to the axis connecting the center of the two windows. Two of them are extended via vacuum pipes to a BaF$_2$ window 40 cm from the chamber center, a mid-infrared radiation thermometer (MIR) is installed. The MIR has a nearly constant sensitivity in the 0.21 - 0.41 eV range (3 - 5.5 μm). It is used here as a radiation intensity meter rather than a thermometer. Although it does not provide spectral resolution, it offers two advantages: it covers a relatively wide energy region near the maximum radiation intensity and has a fast time response (< 1 sec). Thus, in long-term measurements, it is particularly useful for monitoring radiation power from both sides of the sample simultaneously at intervals as short as 1 sec.

Furthermore, X-rays are detected by a silicon drift detector (SDD, Amptek XR-100) positioned outside a Be window, which can measure X-rays down to 1 keV.

Among all instruments, the FT-IR spectrometer plays the central role in quantitative determination of radiant power, since it covers most of the broad peak region and enables model-free estimation of heat flow. The NIR and VL spectrometers complement this by providing spectral information at higher photon energies, which is essential for deducing the radiation temperature. As shown in Fig. 1, when the FT-IR faces the side-A sample, the NIR/VL spectrometers face the side-B sample. To obtain the whole spectrum, we first measure side A with FT-IR (and side B with NIR/VL), then rotate the sample 180° and repeat the measurement. The spectra from both sides are combined and treated as the total emission from the sample.

Calibration of the FT-IR, NIR, and VL spectrometers was performed using radiation from carbon nanotubes, which act as blackbody radiators. Details are provided in the Appendix.

## 3. Experimental procedure

Two types of samples were investigated in the present study:
(1) NiCu sample: Six bilayers of Cu (3.8 nm) and Ni (20 nm) were alternately deposited on a Ni substrate by magnetron sputtering.
(2) pure Ni sample: The Ni substrate used for (1), consisting of 99.9%-purity Ni foil, 0.1 mm thick, with an area of 25 × 25 mm². Details of the NiCu sample preparation are provided in Ref. [5,6].

Each sample thin film was attached to both sides (A and B) of a ceramic heater equipped with an R-type thermocouple. On each side, a 0.3-mm-thick Photoveel plate was placed between the heater and the sample thin film. This was further covered by two Photoveel plates (35 × 35 mm) with a central 20-mm square cutout, and fixed to a holder frame suspended from the chamber lid. Heater power and thermocouple signals were supplied through airtight feedthroughs.

After mounting, the sample was baked in vacuum for three days at ~1150 K and at pressures below 3 × 10$^{-6}$ Pa.

Base line measurements:

Before introducing H$_2$ gas, radiation from the sample was measured to obtain baseline spectra without anomalous heat generation. The heater input voltage was varied as $V_{in}$ = 36, 38, 40, 41, and 42 V. Because equilibrium between input power and radiated power required some time, spectra were recorded 1.0 h and at least 4.0 h after setting each voltage. The measurements were repeated for two cycles of $V_{in}$. These data served to calibrate the radiative power corresponding to zero excess heat.

Excess power measurements:

Radiant power during hydrogen desorption was then measured while evacuating the chamber, i.e., during release of absorbed H$_2$ from the sample. The following sequence was repeated three times for the NiCu sample and twice for the pure Ni sample at $V_{in}$ = 38, 40, and 42 V:
(1) Hydrogen absorption: The chamber was filled with H$_2$ gas to 200–300 Pa, and $V_{in}$ was set to 28 V, giving a heater temperature Tc ≈ 570 K. This condition was maintained for 12–15 h (overnight) without replenishing H$_2$ gas.
(2) Desorption and radiation measurement: The chamber was evacuated while Vin was set to the target value. Full spectra were measured at ~0.5, 2.5, 4.0, and 6.0 h after evacuation. During these intervals, FT-IR continuously monitored the side-A sample, while NIR/VL monitored the side-B sample, with measurements taken every 5 min.

Long-term measurement:

After completing the above cycles, a long-term measurement was performed for the NiCu sample at $V_{in}$ = 40 V without hydrogen refilling. In this run, full spectra were not acquired; instead, both side-A and side-B configurations were fixed, and data were recorded for over 200 h. Each detector acquired data every 5 min.

Throughout all measurements, a data logger continuously recorded (at 1 sec intervals) the following experimental parameters: heater temperature $T_c$, MIR outputs from sides A and B, heater voltage $V_{in}$, heater current $I_{in}$, chamber pressure, and outer wall temperature.

## 4. Results and Discussion

### 4.1. Radiant spectrum

Figure 2 presents a typical example of thermal radiation emitted from the NiCu sample during $H_2$ desorption.

Figure 2(a) shows the whole spectrum obtained approximately 6 hours after evacuation at $V_{in}$ = 40 V, with all detector data overlaid. The blue dots represent FT-IR data, while the red dots denote results from the other detectors. The spectral region between 0.2 and 0.8 eV is covered by both FT-IR and MIR/NIR, and the excellent agreement between them is evident from the near-complete overlap. The red curve corresponds to a gray-body fit to the red data points. Below 0.2 eV, however, the observed radiation clearly exceeds the gray-body prediction—a region not accessible in our previous measurements. This result demonstrates that the total radiation power estimated by the gray-body approximation in Ref. [7] must be revised. In the following analyses, the radiation power is evaluated by integrating the experimental FT-IR spectra over the energy range indicated by the arrows in Fig. 2(a).

Figure 2(b) plots the integrated radiation power in this range as a function of time. Since the FT-IR monitors only the A-side sample, the values correspond to approximately half of the total emission. Because the majority of the radiation originates from the heater, the gradual rise reflects the equilibration process between heater input and radiative output. The increase is well described by a time constant of about 0.7 h.

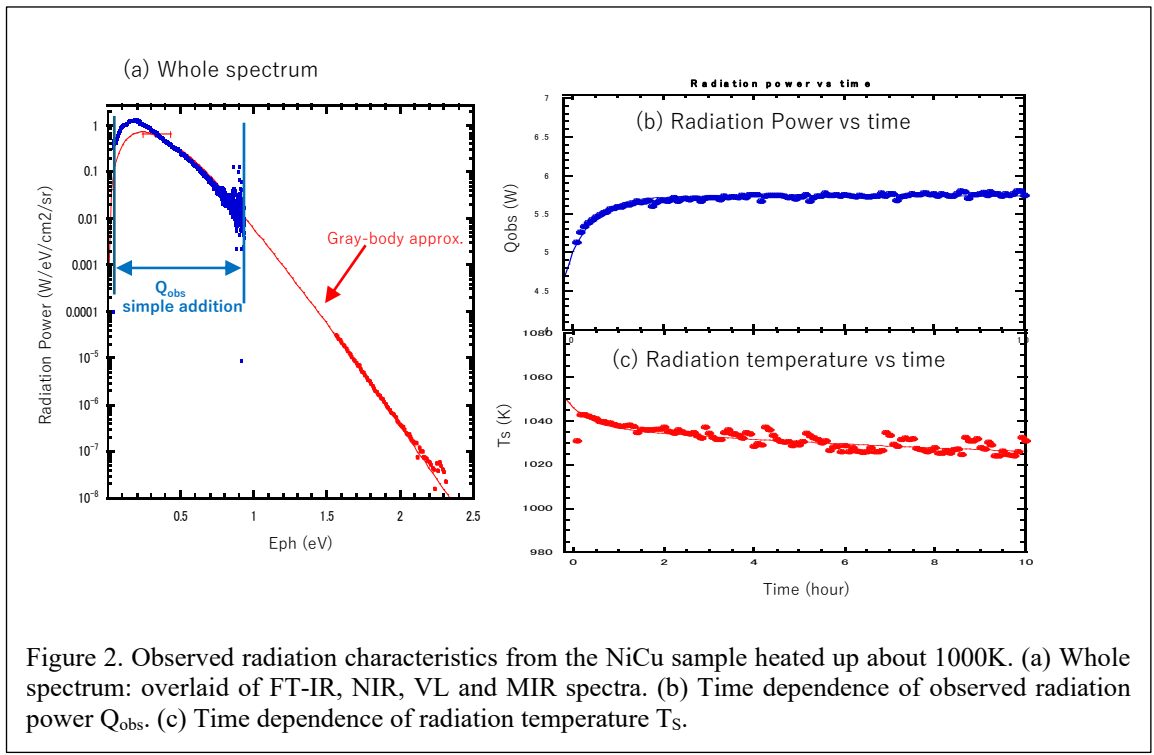

Figure 2. Observed radiation characteristics from the NiCu sample heated up about 1000K. (a) Whole spectrum: overlaid of FT-IR, NIR, VL and MIR spectra. (b) Time dependence of observed radiation power $Q_{obs}$. (c) Time dependence of radiation temperature $T_S$.

Figure 2(c) shows the radiation temperature $T_S$, obtained by fitting the red data in Fig. 2(a) with the gray-body model, as a function of time. Unlike the radiation intensity, $T_S$ rises rapidly and then decreases slowly. Two decay constants, a fast component (~3 h) and a slow component (~80 h), are required to reproduce the observed behavior.

The contrasting time dependences of radiative power and sample temperature indicate that the gray-body assumption, which relates radiation power to the fourth power of temperature, does not hold. The heater temperature $T_C$ exceeds $T_s$ by ~100 K and shows a similar time dependence, suggesting that variations in heat flow cannot be captured solely by monitoring either $T_C$ or $T_s$.

### 4.2. Comparison of spectra with and without $H_2$

Figures 3 and 4 compare radiation spectra obtained with and without hydrogen for the NiCu and pure Ni samples, respectively, both measured at $V_{in}$ = 40 V.

In Figs. 3(a) and 4(a), the black circles denote spectra without $H_2$, and the red circles those with $H_2$. Black and red curves represent gray-body fits to the MIR, FIR, and VL data. The derived $T_s$ values are 1020 K (NiCu) and 1020 K (Ni) with $H_2$, compared with 990 K (NiCu) and 1010 K (Ni) without $H_2$. Thus, both samples exhibit higher radiation temperatures in the presence of hydrogen, with the increase $\Delta T$ being larger for NiCu ($\Delta T$ = 30 K) than for pure Ni ($\Delta T$ = 10 K).

Figs. 3(b) and 4(b) compare the FT-IR spectra directly. As noted in Sec. 4.1, the gray-body approximation reproduces the experimental data above 0.5 eV but underestimates the radiation below 0.4 eV. The observed intensity peaks at 0.15–0.2 eV, whereas the gray-body model predicts a maximum around 0.25 eV. In NiCu (Fig. 3), the spectrum with $H_2$ is clearly enhanced relative to that without $H_2$ for photon energy larger than 0.4 eV, but no enhancement below 0.3 eV.

Figs. 3(c) and 4(c) display the ratio of the spectra with $H_2$ to those without $H_2$. Red circles show the experimental results averaged over 0.024 eV bins to reduce statistical noise, and the blue lines are gray-body ratios calculated for the corresponding temperature increases. For NiCu, the ratio remains near unity below 0.2 eV, but increases rapidly between 0.2 and 0.4 eV. The gray-body model reproduces the data only above 0.4 eV. Since radiation above 0.5 eV contributes less than 10% of the total power, this discrepancy indicates that the gray-body model overestimates the total emission. In contrast, for pure Ni, the ratio increases monotonically with energy, consistent with a modest 10 K temperature rise predicted by the gray-body model.

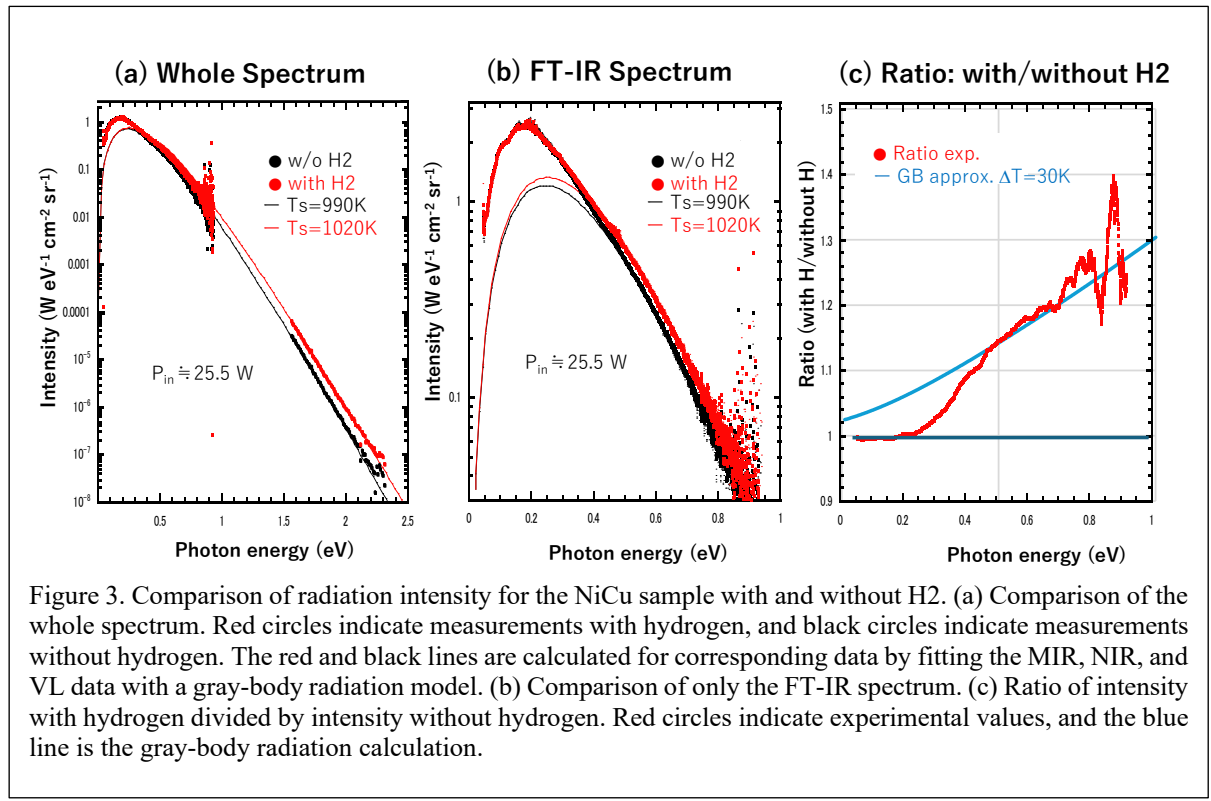

Figure 3. Comparison of radiation intensity for the NiCu sample with and without H2. (a) Comparison of the whole spectrum. Red circles indicate measurements with hydrogen, and black circles indicate measurements without hydrogen. The red and black lines are calculated for corresponding data by fitting the MIR, NIR, and VL data with a gray-body radiation model. (b) Comparison of only the FT-IR spectrum. (c) Ratio of intensity with hydrogen divided by intensity without hydrogen. Red circles indicate experimental values, and the blue line is the gray-body radiation calculation.

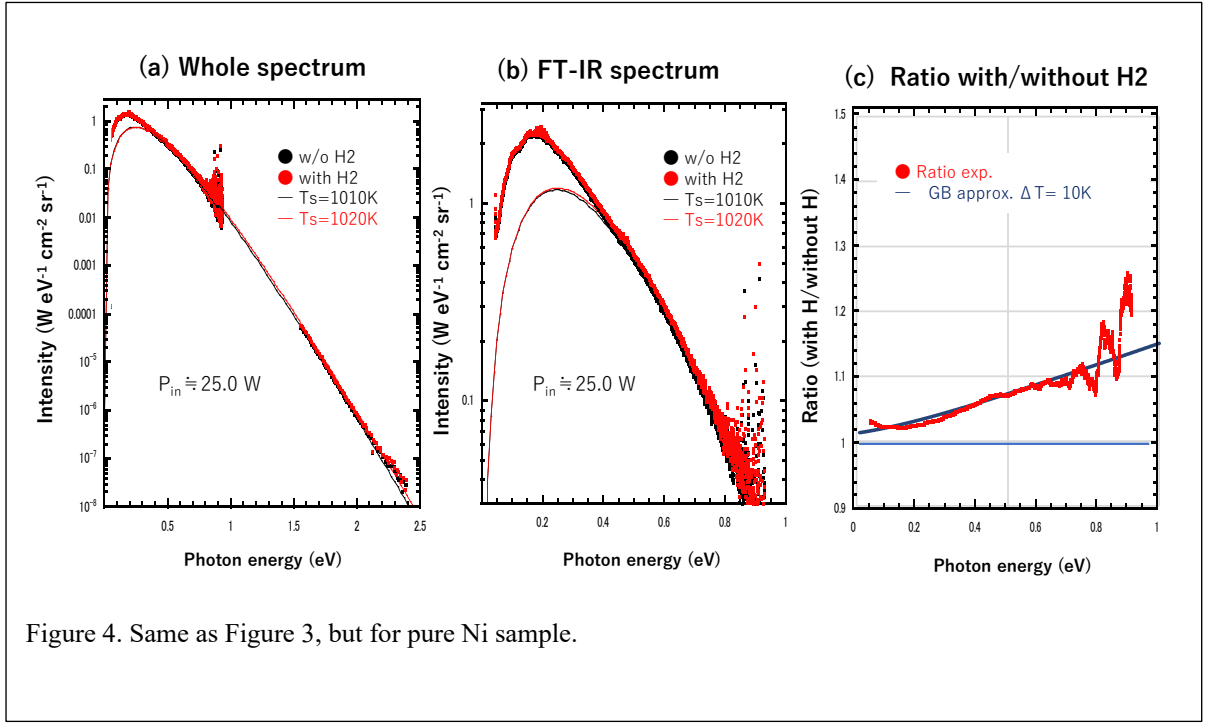

Figure 4. Same as Figure 3, but for pure Ni sample.

The observed radiation $Q_{obs}$ is defined as the FT-IR spectral integral from 0.05 to 0.85 eV multiplied by the angular factor $\pi/2$. For NiCu, the ratio of $Q_{obs}$ with and without $H_2$ is ~1.04, reflecting strong suppression of radiation below 0.4 eV. For pure Ni, where such suppression is absent, the ratio is larger. These results highlight that quantitative heat-flow evaluation without spectral information can be misleading when the radiative properties change upon hydrogen absorption.

### 4.3. Evaluation of excess power

The excess thermal power is evaluated from $Q_{obs}$ using the heat-flow model of Ref. [7]. Two heat sources are considered: heater input $P_{in}$ and sample-generated excess heat $Q_{ex}$. Heat from the heater flows through the insulating spacer (thermal resistance $R_C$) to the sample, where it combines with $Q_{ex}$ to yield the total heat flow $Q_{tot}$. Most of $Q_{tot}$ is dissipated as radiation from the sample ($Q_{obs}$) and from the holder ($Q_H$), connected through another resistance $R_{SH}$. Conductive losses through the supporting rod and reflected radiation are negligible (<0.08 W). At thermal equilibrium:

$$Q_{tot} = P_{in} + Q_{ex} = Q_{obs} + Q_H = Q_{obs}\left(1 + \frac{Q_H}{Q_{obs}}\right). \qquad (1)$$

This shows that $Q_{tot}$ can be deduced from $Q_{obs}$, if $Q_H / Q_{obs}$ is known. The measurement without $H_2$ can give the relationship between $Q_H$ and $Q_S$ because of null $Q_{ex}$.

Figure 5(a) shows $Q_H$ as a function of $Q_{obs}$, where values of $Q_H$ are obtained from $P_{in} - Q_{obs}$. Blue circles represent the data for the NiCu sample, and the black circles represent the data for the pure Ni sample. It seems that $Q_H$ increases at a constant rate as $Q_{obs}$ increases. Blue lines in Fig. 5(a) are the results of applying a linear function, $Q_H = aQ_s+b$, to each data set. The best fit gives a = 3.584 and b = 0.933 for the NiCu sample and a = 2.807 and b = 1.071 for the pure Ni sample. However, this form does not satisfy the condition $Q_H = 0$ at $Q_s = 0$.

Red lines are the best fits when applying the power function, $Q_H = A\,Q_S^m$, which is effective as an approximation function for $Q_H$ for the thermal radiation intensity being proportional to the fourth power of the temperature. The linearity in the measured region is reproduced with the exponential parameter m being $0.72 < m < 0.76$ for the NiCu sample and $0.77 < m < 0.81$ for the pure Ni sample. Nevertheless, within the observed range, the linear approximation is sufficiently accurate and was adopted for further analysis:

$$Q_{tot} = \alpha + \beta Q_{obs} \quad (2)$$
$$Q_{ex} = Q_{tot} - P_{in} \quad (3)$$

with α = 3.584 and β = 1.933 for the NiCu sample and α = 2.807 and β = 2.071 for the pure Ni sample.

Control tests confirmed that the holder material (Photoveel) does not absorb hydrogen, and its radiation spectrum remained unchanged (<0.2% difference). Therefore, Eqs. (2) and (3) were used to determine $Q_{ex}$.

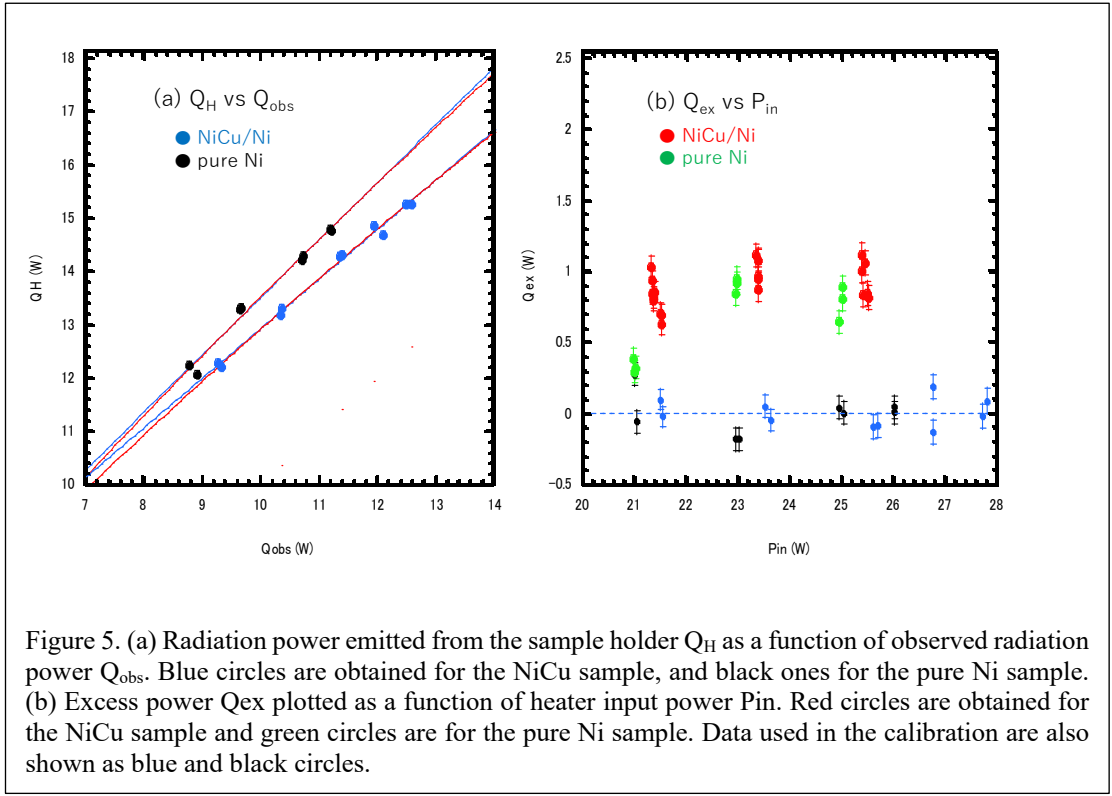

Figure 5. (a) Radiation power emitted from the sample holder $Q_H$ as a function of observed radiation power $Q_{obs}$. Blue circles are obtained for the NiCu sample, and black ones for the pure Ni sample. (b) Excess power Qex plotted as a function of heater input power Pin. Red circles are obtained for the NiCu sample and green circles are for the pure Ni sample. Data used in the calibration are also shown as blue and black circles.

Figure 5(b) shows $Q_{ex}$ as a function of $P_{in}$. Red and green circles denote data with $H_2$ for the NiCu and pure Ni sample, respectively, while blue and black circles denote the corresponding data without $H_2$. All values were obtained at least four hours after voltage setting, ensuring thermal equilibrium. The typical uncertainty in $Q_{ex}$ is ~0.08 W (1σ). The results clearly show excess power of ~1 W: Qex ≈ 1.1 W for the NiC sample at $P_{in}$>21 W and just under 1 W for the pure Ni sample at $P_{in}$ > 23 W.

Compared with our earlier report [7], two important differences emerge: (i) the excess power for NiCu is much smaller than previously reported, and (ii) a clear excess is observed for pure Ni. While sample-to-sample variation cannot be excluded, the present inclusion of spectral data below 0.5 eV provides a more accurate evaluation of excess heat. In NiCu, radiation below 0.3 eV is not enhanced by the presence of $H_2$, whereas for Ni no such suppression is observed.

### 4.4. Long term measurement

Figure 6 shows long-term data for NiCu acquired at $V_{in}$ = 40 V over 215 h, with spectra collected every 5 min without replenishing hydrogen. Because only one side of the sample was monitored, the $Q_{ex}$ values exhibit larger fluctuations than in Fig. 5(b).

The radiation rises to ~1.2 W after ~10 h and then decays gradually—initially faster, and more slowly after ~150 h. We attribute this decrease to depletion of hydrogen fuel within the sample. The time dependence is expressed as:

$$Q_{ex}(t) = (a - b \exp\left(-\frac{t}{\tau_{eq}}\right)) \times (A_F \exp\left(-\frac{t}{\tau_F}\right) + A_S \exp\left(-\frac{t}{\tau_S}\right))^2. \quad (4)$$

The first bracket corresponds to the increase in radiation intensity up to thermal equilibrium, with its time constant $\tau_{eq}$. The second bracket corresponds to the decrease in H density, with the fast decay time constant $\tau_F$ and the slow decay time constant $\tau_S$. Here, we assume that two H atoms are involved in a reaction, and, thus, the reaction rate is the square of the density (although in this case the exponential decay function is an approximation).

The red curve in Fig. 6 shows a fit with a=0.93, b=0.46, $\tau_{eq}$=5.0, $A_F$=0.2, $\tau_F$=90, $A_S$=1.0, and $\tau_S$=2500. The fit reproduces the data well, and extrapolation suggests energy production could continue for ~4000 h until hydrogen is exhausted. The integrated output is 4.26 MJ. The absorbed hydrogen was measured prior to the run, with an upper limit of $1\times10^{-4}$ mol. Assuming full participation, the minimum specific energy is E/H > 22 keV per hydrogen atom—orders of magnitude greater than any chemical process. The actual value must be much higher, as most absorbed hydrogen is lost by thermal diffusion and thus cannot contribute to the reaction.

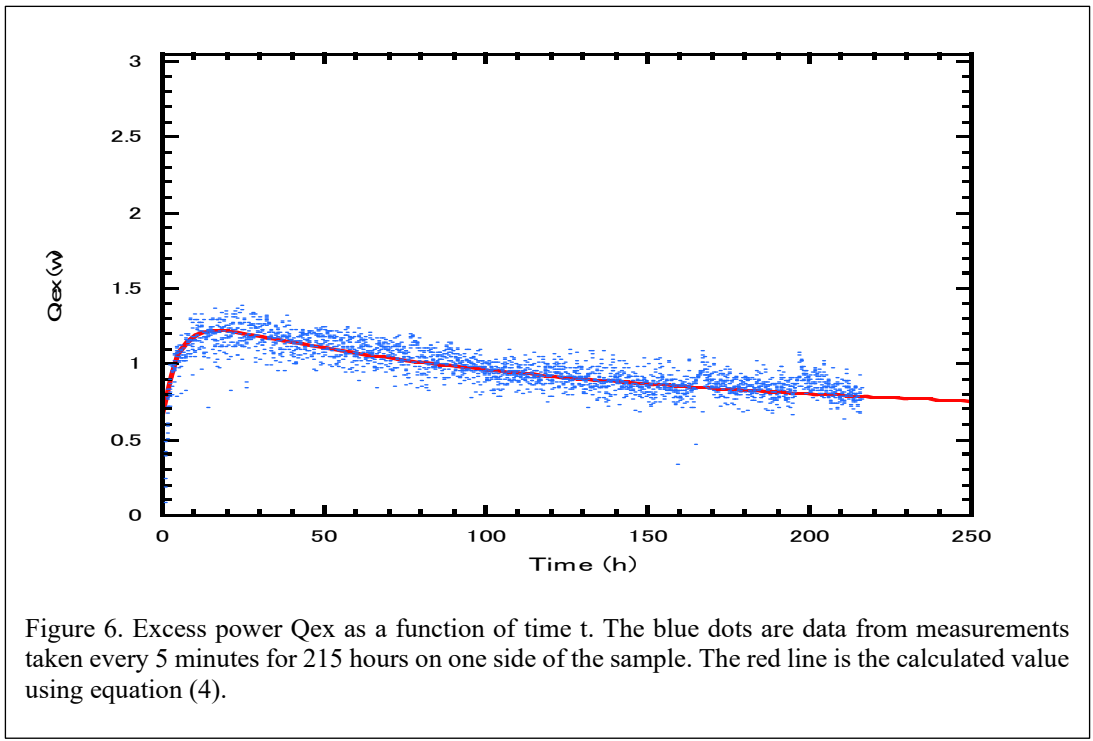

Figure 6. Excess power Qex as a function of time t. The blue dots are data from measurements taken every 5 minutes for 215 hours on one side of the sample. The red line is the calculated value using equation (4).

## 5. Conclusions

The present experiment, which expanded the energy range for spectroscopic measurement of radiation, revealed the following findings:

Thermal radiation spectra of both the NiCu composite and pure Ni samples deviate significantly from that of simple gray-body radiation. Unlike the gray-body calculation, where intensity decreases below 0.3 eV, the observed spectra exhibited a continued rise, with a peak at around 0.2 eV.

In the presence of hydrogen, both samples showed an increase in radiation power by several percent, but with distinct spectral features. For the NiCu sample, little enhancement was observed below 0.3 eV, whereas a sharp increase appeared above 0.35 eV. In contrast, the pure Ni sample exhibited a gradual enhancement starting from as low as 0.05 eV.

The total radiation output was determined directly from the observed spectra, without reliance on a gray-body radiation model. The excess heat generation in the NiCu sample was estimated to be approximately 1.1 W, while the pure Ni sample exhibited a slightly lower value.

Long-term measurements over 215 hours demonstrated that excess power generation can persist for extended periods, with a decay time exceeding 2000 hours. The energy released per absorbed hydrogen atom was estimated to exceed 22 keV.

While these results strongly support the occurrence of anomalous heat generation, the measurement of excess energy alone is insufficient to establish whether the origin lies in a novel nuclear process. To provide conclusive evidence, signatures of nuclear reactions must be identified, such as residual nuclei, nuclear γ-rays, high-energy charged particles, or characteristic X-rays from sample atoms. The advantage of the present experimental system lies in its capability to simultaneously monitor various radiations together with excess heat, offering a powerful tool for further investigation into the nature of this anomalous phenomenon.

**Acknowledgements**
This work is supported by Clean Planet Inc., and The Thermal & Electric Energy Technology Foundation.

**Appendix**
The intensity calibration of the three spectrometers, FT-IR, NIR and VL, was performed by setting two sheets of carbon nanotube (CNT) at the sample positions. Since the emissivity of CNT is more than 0.98, the radiation is treated as blackbody radiation; the error due to this is less than 0.3% in the analysis. The blackbody radiation power $B_{black}$ is expressed for the photon with the energy $E_{ph}$ emitted from the surface with the temperature of $T$ as

$$B_{black}(E_{ph}, T) = \frac{a_1 E_{ph}^3}{\exp\left(a_2 \frac{E_{ph}}{T}\right) - 1} \quad [\text{W cm}^{-2} \text{ sr}^{-1} \text{ eV}^{-1}], \quad (A1)$$

where $a_1 = 5.07 \times 10^3$ [W cm$^{-2}$ sr$^{-1}$ eV$^{-4}$] and $a_2 = 1.16 \times 10^4$ [K eV$^{-1}$].

In the CNT measurement, $T$ is measured as $T_{MIR}$, the temperature output of the MIR detector for the emissivity 1.0. Using this temperature, the absolute intensity of the radiant power is calculated and compared with the measured raw spectrum to obtain a correction factor for each photon energy. For the detectors, FT-IR, NIR and VL, the correction factor $F(E_{ph})$ is determined as

$$F(E_{ph}) = \frac{B_{black}(E_{ph}, T)}{Y_{raw}(E_{ph}, T)}, \quad (A2)$$

where $Y_{raw}(E_{ph}, T)$ is a raw yield measured for CNT with temperature $T$. Since the correction should depends only on $E_{ph}$, the correction factor finally used is an average of several measurements taken at different temperatures. The correction includes all the effects; absorption due to the window material and fiber cable, the detection efficiency of the photosensor, and the solid angle of the detection.

As an example to calibrate the FT-IR detector, we show two spectra the raw data at two different temperatures, the calibration curve obtained using data at five different temperatures, and the calibrated spectrum.

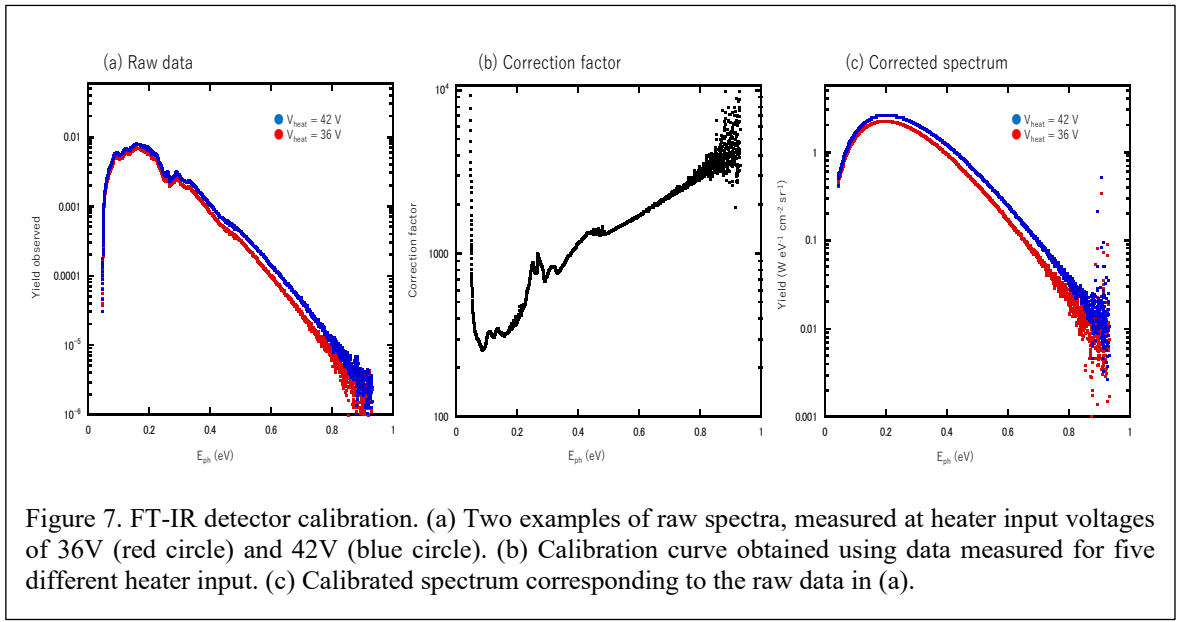

Figure 7. FT-IR detector calibration. (a) Two examples of raw spectra, measured at heater input voltages of 36V (red circle) and 42V (blue circle). (b) Calibration curve obtained using data measured for five different heater input. (c) Calibrated spectrum corresponding to the raw data in (a).

References
[1] M. Fleischmann and S. Pons, J. Electroanal. Chem. 261 (1989) 301.
[2] E. Storms, Naturwissenshaften, 97, 861 (2010).
[3] L.O. Freire, D.A.d. Andrade, J. Electroanal. Chem. 903, 115871 (2021).
[4] A. Kitamura, et al., Phys. Lett. A373 (2009) 3109; Int. J. of Hydrogen Energy 43 (2018) 16187.
[5] Y. Iwamura, et al., J. Condensed Matter Nucl. Sci. 33 (2020) 1; ibid., 36 (2022) 285.
[6] T. Itoh et al, J. Condensed Matter Nucl. Sci. 36 (2022) 274.
[7] J. Kasagi et al., arXive: 2311.18347, J. Condensed Matter Nucl. Sci. 39 (2025) 210–219.

---

[i] # Paper presented at International Conference, ICCF26.